\documentclass[showpacs, twocolumn]{revtex4-1}
\usepackage{graphicx}
\usepackage{amsmath}
\usepackage{appendix}

\begin{document}

\title{ Self-consistent theory of Bose-Einstein condensate with impurity at finite temperature}

\author{Abdel\^{a}ali Boudjem\^{a}a}

\affiliation{Department of Physics, Faculty of Sciences, Hassiba Benbouali University of Chlef P.O. Box 151, 02000, Ouled Fares, Chlef, Algeria.}

\email {a.boudjemaa@univ-chlef.dz}


\begin{abstract}
We study the properties of Bose-Einstein condensate (BEC)-impurity mixtures at finite temperature
employing the Balian-V\'en\'eroni (BV) variational principle. The method leads to a set of coupled nonlinear equations of motion for the condensate and its 
normal and anomalous fluctuations on the one hand, and for the impurity on the other. We show that the obtained equations satisfy the energy and number conserving
laws.
Useful analytic expressions for the chemical potential and the radius of both condensate and anomalous components are derived
in the framework of the Thomas-Fermi (TF) approximation in $d$-dimensional regime. Effects of the impurity on these quantities are discussed.

\end{abstract}


\maketitle

\section{Introduction} \label{Intro}

In the past decade, BEC-impurity mixtures have been the subject of intense experimental \cite {Chik, Ciam, Gunter, Bew, Ospe, Pal, Brud, Will, Cat, bloch} 
and theoretical  \cite {Jack1, Tim1, Tim2, Tim3, Tim4,Temp, Jack2, Jack3, Tim4,Tim5, Stef} studies.  
These mixtures have been also investigated in the case where the surrounding medium is fermionic atoms, 
the so-called Fermi polaron problem (see for review \cite {Piet}).

Most of the above studies ignore completely the behavior of BEC-impurity systems at finite temeprature. 
The effects of the temperature are however important, in particular on the fluctuations, on the expansion of the condensate, and on the thermodynamics of the system.
Certainly, the dynamics of the BEC-impurity  at nonzero temperature is a challenging problem since for example the Bogoliubov approximation becomes invalid, at least at large times, and large
thermal phase fluctuations have to be taken into account even at low temperatures where density fluctuations are small.
It is therefore instructive to derive a self-consisent approach describing the static and the dynamics of BEC-impurity mixtures
at finite temperature especially since that all experiments actually take place at nonzero temperatures.

Our approach is based on the time-dependent BV  variational principle \cite {BV1, BV2, Bon, koo, BV3, BV4, BV5}.  
During the last three decades, this principle has been applied in different area of physics. First, it has been used to various quantum problems including heavy ion reactions \cite {BV1, BV2, Bon, koo}, quantum fields out of equilibrium \cite{Jakiw, Cic} attempts to go beyond the Gaussian approximation for fermion systems \cite {Flu}. Furthermore, the BV variational principle has been employed to provide the best approximation to the generating functional for multi-time correlation functions of a system of bosonic and fermionic observables \cite {BV3, BV4, BV5, Cic1, Ben1}. Recently, it has been applied to derive a set of equations governing the dynamics of trapped Bose gases \cite {Ben2,Ben3, boudj2010, boudj2011}. The BV variational principle is also used to determine the particle number fluctuation in fragments of many-body systems of fermions \cite{Ced}.

The time-dependent BV variational principle requires first the choice of a trial density operator.
In our case, we will consider a Gaussian time-dependent density operator. This ansatz, which belongs to the class
of the generalized coherent states, allows us to perform the calculations since there exists Wick’s theorem, while retaining
some fundamental aspects such as the pairing between atoms. The BV variational principle is based on the minimization of an action 
which involves two variational objects : one related to the observables of interest
and the other is akin to a density matrix (see below). This leads to a set of coupled time-dependent mean-field equations 
for the condensate, the noncondensate, the anomalous average and the impurity. We call this approach “time-dependent Hartree-Fock-Bogoliubov” (TDHFB). 

The rest of the paper is organized as follows: In Sec.\ref{BV}, we review the main features of the BV variational principle which we use to derive the TDHFB equations.
In Sec.\ref{applic}, the TDHFB equations are applied to a trapped BEC-impurity system to derive
coupled equations of motion for the condensate, the noncondensate, the anomalous density and for the impurity. The link between our equations and those existing in the literature 
such as the Hartree-Fock-Bogoliubov-de-Gennes (HF-BdG) theory is highlighted. 
Sec.\ref{Hydr} is devoted to discuss the hydrodynamic approximation and conservation laws. 
Moreover, useful expressions for the radius and the chemical potential of BEC-impurity and anomalous density-impurity mixtures 
are obtained in the TF limit in $d$-dimensional case. 
Our conclusion and outlook are presented in Sec.\ref{concl}.

\section { Balian- V\'en\'eroni variational principle and TDHFB equations} \label{BV}

The BV variational principle \cite {BV3, BV4} is based on the Liouville-Von Neumann equation. It addresses the following question:
Given the state of a system described by the density operator $D (t_i)$ at an initial time $t_i$, what will be the expectation value of an observable $A$ measured at a final time $t_f$ ?
The answer can be stated as follows: take two variational objects, one for an observable denoted  ${\cal A}$ and one for a density operator denoted ${\cal D}$,
 the optimal value of $\text{Tr} A D ({t_f})$ is given by the stationary value of the action-like functional \cite {Cic, Ben2}
\begin{equation}  \label{action}
I=\text{Tr} A {\cal D} ({t_f})-\int_{t_i}^{t_f} dt \text{Tr} {\cal A} \left( \frac{d{\cal D}}{dt}+i[H,{\cal D}]\right),
\end{equation}
with the two mixed boundary conditions:
\begin {align}
{\cal D}(t_i)= D (t_i)  \qquad {\cal A}(t_f)=A. 
\end {align}
The symbol $\text{Tr}$ stands for a trace taken over a complete basis of the Fock space, and $H$ is the Hamiltonian of the system assumed to be time independent.
The variation of  $I$ in (\ref{action}) provides evolution equations for the operators ${\cal D}$ and ${\cal A}$, which are in general coupled. \\
When the allowed variation  $\delta {\cal A} (t)$ and $\delta {\cal D} (t)$ are unrestricted, the stationary conditions of the BV functional (\ref{action}) 
lead to the exact Liouville-von Neumann equation for$ {\cal A} (t)$ and the (backward) Heisenberg equation for ${\cal D} (t)$ (see \cite {BV3} for more details). 

Before introducing the trial class of operators that we shall use in this work for $ {\cal A} (t)$ and ${\cal D} (t)$, it is useful to define the $2n$-component operator $\alpha$  \cite{ Ben2, boudj2011}
\begin{equation}  \label{ope}
\alpha=\left(
\begin{array}{c}
 \psi \\
 \psi^+
\end{array}
\right),
\end{equation}
where  $\psi$  and $ \psi^+$ denote boson field creation and annihilation operators in a given single-particle space of dimension $n$. 
The commutation relations can be expressed in the compact form  $(i,j=1\cdots 2n)$
\begin{equation}  \label{comm}
[\alpha_i,\alpha_j]=\tau_{ij},
 \end{equation}
where $\tau=i\sigma_2$ is $(2n\times 2n)$ second Pauli matrix.
With this notation, the trial class for the operator ${\cal D} (t)$, the exponentials of the linear plus quadratic forms in $\alpha$,
can be written in the factorized form \cite{Ben2, BB66, BB69}
\begin{equation}\label{eq1}
{\cal D} (t) = \exp{(\nu )} \exp{(\tilde {\lambda} (t)\tau\alpha )} \exp{\left(\frac{1}{2}\tilde {\alpha}\tau S (t)\alpha \right)},
 \end{equation}
where $\nu$  is a  c-number, $\lambda $ is a $2n$ -component  vector  and $S$ is  a $(2n\times 2n)$ symplectic matrix 
($\tau S$ is symmetric). The tilde symbol denotes vector or matrix transposition. More detailed properties of the exponential operator (\ref{eq1}) cand be found in \cite{BB66, BB69}.\\
For any operator $O$, let us denote the average value$\langle O\rangle=\hbox{Tr} (O {\cal D}) /{\cal Z} $ and the shifted operator 
\begin{equation} \label{SO}
\bar O=O-\langle O\rangle.
\end{equation} 
Then, $ {\cal D}$ is completely specified by the knowledge of “the partition function” ${\cal Z}(t)=\hbox{Tr}\,{\cal D}(t)$,
the vector $ \langle \alpha \rangle$ and the one-body density matrix $\rho$, are  defined as
\begin{equation} \label{dm}
\rho_{ij}=\langle(\tau\bar{\alpha})_j\bar{\alpha}_i\rangle.
\end{equation} 
It is convenient to characterize the operator $ {\cal D}$ by ${\cal Z}$, $\langle \alpha\rangle$ and $\rho$  instead of $\nu$, $\lambda$ and $S$. 
Indeed, we will see that the evolution equation for the former variables take a simple form.\\
With the same notation,  the variational choice  for the operator $ {\cal A}$ can be taken as a linear plus quadratic form in the operator $\alpha$ \cite {Ben2}:
\begin{equation}\label{eq2}
{\cal A} (t) =\nu +\tilde {\lambda} (t)\tau\alpha + \frac{1}{2}\tilde {\alpha}\tau S (t)\alpha,
 \end{equation}
to which we shall loosely refer as a single-particle operator.\\
To get a reduced form of the functional (\ref{action}),  corresponding to (\ref{eq1}) and (\ref{eq2}), one should calculate the quantities 
$\text{Tr} {\cal A} d{\cal D}/dt$ and $\text{Tr}  {\cal A} [H,\,{\cal D}]$.
The former can be evaluated by calculating first $\text{Tr} {\cal A D}$ using Wick theorem. This yields
\begin{equation}\label{e7}
\text{Tr} {\cal AD}\equiv {\cal Z}\,\langle A\rangle={\cal Z}\left \{\nu+\tilde{\lambda} \tau \langle \alpha\rangle-\frac{1}{2} \text{tr} S (\rho- \langle \alpha\rangle\langle \tilde{\alpha}\rangle\tau)\right \},
\end{equation}
where the symbol tr indicates a trace in the single-particle space of the ($2n\times 2n$) matrices.
Keeping fixed the parameters of ${\cal A}$, differentiation with respect to time gives:
\begin{equation}\label{e8}
\text{Tr} {\cal A} \frac{d{\cal D}}{dt}= \frac{\dot{\cal Z}}{{\cal Z}} \text{Tr} {\cal AD}+{\cal Z}\left \{ \tilde{L}\tau \langle \dot \alpha\rangle-\frac{1}{2} \text{tr} S \dot\rho\right \},
\end{equation}
with ${L}=\lambda-S\langle\alpha\rangle$ and the dots denote the time derivative $d/dt$. \\
The second term of the integrand in the functional (\ref{action}), $\text{Tr}  {\cal A} [H,\,{\cal D}]$, can be
written without specifying the Hamiltonian as
\begin{equation}\label{e9}
\text{Tr} {\cal A} [H,\,{\cal D}]={\cal Z} \left\{ \tilde{L}\tau P- \frac{1}{2} \text{tr} S \left[\rho, M\right] \right\} .
\end{equation}
The expressions for the vector $P$ and the matrix $M$ are the following:
\begin{equation}\label{e10}
P_i=\sum_{j=1}^{2n}\tau_{ij}\frac{\displaystyle \partial {\displaystyle \cal E}}{\displaystyle \partial \langle \alpha\rangle}_j \, , M_{ij}=-2\frac{\displaystyle d{\cal E}}{\displaystyle d\rho_{ji}},
\end{equation}
where ${\cal E}=\langle H\rangle$ is the mean field energy.\\
The matrix $M$ is the analog of the Hartree-Fock Hamiltonian for fermions.
The vector $P$ has no equivalent for fermions.
By means of Wick theorem, the energy ${\cal E}$ can be evaluated in terms of $\langle \alpha\rangle$ and $\rho$ only.
From the definitions (\ref{e10}), one sees that $P$ and $M$ are independent of ${\cal Z}$. Finally, the BV action  takes the form \cite{Cic1, Ben2}:
\begin {widetext} 
\begin{equation}  \label{action1}
I=\text{Tr} A {\cal D} ({t_f})-\int_{t_i}^{t_f} dt \left[\text{Tr}  {\cal A D} \frac{\dot{\cal Z}}{{\cal Z}}-i{\cal Z} \left\{ \tilde{L}\tau \left(i\langle \dot\alpha\rangle-\tau \frac{\partial {\cal E}}{\partial \langle \alpha\rangle}\right)-\frac{1}{2} \text{tr} S \left( i\dot \rho +2\left[\rho,\frac{d{\cal E}}{d\rho}\right]\right)\right\}\right].
\end{equation}
\end {widetext} 
The equation of motion for ${\cal Z}$, $\langle \alpha\rangle$ and $\rho$ are now obtained by writing the stationary of the BV functional (\ref{action1}) 
with respect to $\nu$, $\lambda$ and $S$. These equations can be written in compact form \cite{Cic1, Ben2, boudj2011}
\begin{equation} \label {TDHFB1}
i\dot {\cal Z}=0,
\end{equation}
\begin{equation} \label {TDHFB2}
i\langle \dot \alpha\rangle =\frac{d{\cal E}}{d\langle \alpha\rangle},
\end{equation}
and 
\begin{equation} \label {TDHFB3}
i\dot \rho =-2\left[\rho, \frac{d{\cal E}}{d\rho}\right].
\end{equation}
We see from equation (\ref {TDHFB1}) that the partition function $ {\cal Z}$ is a constant of motion. The two last equations (\ref{TDHFB2}) and (\ref{TDHFB3}) 
imply that the energy ${\cal E}$ is conserved when the Hamiltonian $H$ does not depend explicitly on time. 
We may notice at this point that the above equations constitute a  closed self-consistent system.\\
One of the most noticeable properties of the these equations is the unitary evolution of the one body density matrix $\rho$, which means
that the eigenvalues of $\rho$ are conserved. This leads to the expression \cite{Cic2, boudj2011}
\begin{equation} \label {Invar}
C=4\rho (\rho+1)+1,
\end{equation}
where $C$ known as the Heisenberg invariant.\\
Therefore, Eq.(\ref{Invar}) implies the conservation of the von Neumann entropy $S =-\hbox{Tr}\,{\cal D} \ln{\cal D}$. \\
Other property of the TDHFB equations is that they admit a Lie-Poisson structure \cite {Ben4, Cic2}.
Also, among the advantages of the TDHFB equations is that they should in principle yield the general time, space, and temperature dependence of various densities. 
In addition, the most important feature of the TDHFB equations is that they are valid for any density matrix operator of the form (\ref{eq1}). 

\section{Application to BEC-impurity mixture} \label{applic}
We consider $N_I$ impurity  bosonic atoms of mass $m_I$ in an external trap $V_I({\bf r})$, and identical bosons of mass $m_B$ trapped by an external potential $V_B({\bf r})$. 
The impurity-boson interaction and boson-boson interactions have been approximated by the contact potentials $g_B \delta ({\bf r}-{\bf r'})$ and $g_{IB} \delta ({\bf r}-{\bf r'})$,
respectively, where $g_B=(4\pi \hbar^2/m_B) a_B$ and $g_{IB}=2\pi \hbar^2 (m_B^{-1}+m_I^{-1}) a_{IB}$ with 
$a_B$ and $a_{IB}$ being the boson-boson  and  impurity-boson scattering lengths, respectively. 
We neglect the mutual interactions of impurity atoms since we  assume that their number and local density remains sufficiently small \cite {Tim1, Jack1} and hence there is no impurity fluctuation.
The many-body Hamiltonian for combined system which  describes  bosons, impurity and impurity-boson gas coupling is given by 
\begin{eqnarray} \label{eq4}
&& \hat H =\hat H_B+\hat H_I+\hat H_{IB} \\&& \nonumber
=\int d{\bf r} \hat\psi_B^{+}({\bf r}) \left[-{\frac{\hbar^2}{ 2m_B}}\Delta + V_B({\bf r}) +\frac{g_B}{2} \hat \psi_B^{+}({\bf r})\hat \psi_B({\bf r})\right]\hat\psi_B({\bf r}) \\&& \nonumber
+\int d{\bf r} \hat\psi_I^{+}({\bf r}) \left[-{\frac{\hbar^2}{ 2m_I}}\Delta + V_I({\bf r})\right]\hat\psi_I({\bf r})\\&& \nonumber
+g_{IB}\int d{\bf r} \hat\psi_I^{+}({\bf r}) \hat \psi_I({\bf r})\hat \psi_B^{+} ({\bf r}) \hat\psi_B({\bf r}),
\end{eqnarray}
where  $\hat\psi_B(\bf r)$ and $\hat\psi_I({\bf r})$ are the boson and impurity field operators.\\
Treating this Hamiltonian using a self-consistent quadratic approximation, and expanding the field operators using (\ref{SO}) $\hat\psi=\hat{\bar{\psi}}+\Phi$, where $\hat{\bar{\psi}}$ is the noncondensed part of the field operator, one finds
\begin {widetext} 
\begin{align} \label{ham}
& \hat H  = \int d{\bf r} \left \{\Phi_B^*\left (h_B^{sp}+\frac{g_B}{2}|\Phi_B|^2\right)\Phi_B +\bar{\psi}_B^+ \left(h_B^{sp}+2g_B n\right)\bar{\psi}_B  
+\frac{g_B}{2}\left [(\tilde{m}^{*}+\Phi_B^{*2}) \bar{\psi}_B\bar{\psi}_B +(\tilde{m}+\Phi_B^2)\bar{\psi}_B^+\bar{\psi}_B^+ \right]\right\}  \\  
&+\int d{\bf r} \left(\Phi_I^* h_I^{sp}\Phi_I \right) \nonumber\\ 
&+ g_{IB}\int d{\bf r} \left [ (|\Phi_B|^2+\bar{\psi}_B^+\bar{\psi}_B) |\Phi_I|^2 \right],   \nonumber
\end{align}
\end {widetext}
where $h_B^{sp}=-(\displaystyle\hbar^2/\displaystyle 2m_B) \Delta + V_B$ and $h_I^{sp}=-(\displaystyle\hbar^2/\displaystyle 2m_I)\Delta + V_I$ are, respectively the single particle Hamiltonian for the condensate and the impurity. $\Phi_B$ and $\Phi_I$ stand for the condensate and the impurity wave functions, respectively. The noncondensed density $\tilde{n}$ and the anomalous density $\tilde{m}$ of the condensate are identified, respectively, as $\langle\bar {\psi}_B^{+}\bar {\psi}_B\rangle$, $\langle\bar {\psi}_B\bar {\psi}_B\rangle$, 
where $n=|\Phi_B|^2+\tilde{n}$ is the total density.
The coefficients of linear terms in $\bar{\psi}$,$\bar{\psi}^{+}$ in (\ref {ham})  can be shown to vanish\cite{Griff}.\\
Having defined our notations, we can now easily write down the density matrix (\ref {dm}) 

\begin{equation}
\rho=\begin{pmatrix} 
\langle \bar{\psi}^{+}\bar{\psi}\rangle & -\langle\bar{\psi}\bar{\psi}\rangle\\
\langle\bar{\psi}^{+}\bar{\psi}^{+}\rangle& -\langle\bar{\psi}\bar{\psi}^{+}\rangle
\end{pmatrix}
=\begin{pmatrix} 
\tilde{n} & -\tilde{m} \\
\tilde{m}^{*}& -(\tilde{n}+1)
\end{pmatrix}.
\end{equation} 
The coupling between different quantities of the system occurs via the derivatives of ${\cal E}=\langle \hat H\rangle$ which can be calculated in usual manner.\\
Making use of these expressions, Eqs.(\ref {TDHFB2}) and (\ref {TDHFB3}) take the following explicit form:
\begin {widetext} 
\begin{subequations}\label{E:gp}
\begin{align} 
& i\hbar \dot{\Phi}_B  = \left[ h_B^{sp}+g_B (|\Phi_B|^2+2\tilde{n}) + g_{IB} |\Phi_I|^2 \right]\Phi_B + g_B\tilde{m}
\Phi_B^{*} ,  \label{E:gp1} \\  
&i\hbar \dot{\Phi}_I  = \left[ h_I^{sp} +g_{IB} (|\Phi_B|^2+\tilde{n})\right]\Phi_I ,  \label{E:gp2} \\ 
&i\hbar \dot{\tilde{n}}  = g_B\left(\tilde{m}^{*}\Phi_B^2-\tilde{m} {\Phi_B^{*}}^2\right) ,  \label{E:gp3} \\ 
&i\hbar \dot{\tilde{m}} = g_B (2\tilde{n} +1)\Phi_B^2 + 4\left[ h_B^{sp}+2g_B n+{\frac{\displaystyle g_B}{\displaystyle 4}}(2\tilde{n} +1)+g_{IB} |\Phi_I|^2\right]\tilde{m}   \label{E:gp5}.
\end{align}
\end{subequations}
\end {widetext}
Putting $g_{IB} =0$ (i.e., neglecting the mean-field interaction energy between bosons and impurity components)
one recovers the usual TDHFB equations \cite {boudj2010, boudj2011, boudj2012, boudj 2013} describing a degenerate Bose gas at finite temperature
and the Schr\"o\-dinger equation describing a noninteracting impurity system. 
For $\tilde{n}=0$ and $\tilde{m}=0$, the TDHFB equations reduce to the well known Gross-Pitaevskii equation.
One should mention here that Eqs.(\ref{E:gp1}) and (\ref{E:gp2}) are more general than those used in the literature \cite {Jack1, Tim1, Tim2, Tim3, Tim4,Temp, Jack2, Jack3, Tim4,Tim5, Stef}
since they contain simultaneously the impurity and the condensate fluctuations. 
Equation. (\ref{E:gp5}), which describes the behavior of the anomalous density-impurity, has never been derived before in the literature.

For impurities in a fermionic bath, evidently Eq.(\ref{E:gp1}) has no analog, the corresponding equations of (\ref{E:gp3}) and (\ref{E:gp5}) are the Hartree-Fock and the gap equations, respectively.
While Eq.(\ref{E:gp2}) becomes a set of Schr\"o\-dinger equations describing a noninteracting Fermi system with $\Phi_I=\sum_j {\Phi_I^{(j)}}$.

A useful relation between the normal  and anomalous densities can be given via Eq.(\ref {Invar}):
\begin{equation}  \label{eq6}
\begin{array}{rl}
C_B& =(2\tilde{n} +1)^2-4|\tilde{m} |^2.
\end{array}
\end{equation}
In the uniform case,  by working in the momentum space, $C_k=\coth^2 ( \varepsilon_k/T)$ \cite {boudj2012},  where $\varepsilon_k$ is the excitation energy of either BEC or impurity.
At zero temperature $C=1$. 
The expression of $C$ allows us to calculate in a very useful way the dissipated heat for $d$-dimensional Bose gas as 
$Q=(1/n)\int E_k C_k d^dk/(2\pi)^d$ with $ E_k=\hbar^2 k^2/2m$ \cite {boudj2012}.
Indeed, the dissipated heat or the superfluid fraction are defined through the dispersion of the total momentum operator of the whole system. This definition is valid for an arbitrary system, including nonequilibrium one. In an equilibrium system, the average total momentum is zero. Hence, the corresponding heat becomes just the average total kinetic energy per particle. \\
It is necessary to stress that our formalism provides an interesting formula for the superfluid fraction $n_s=1-2Q/d T$ in $d$-dimensional Bose gas \cite {boudj2012, Yuk} which reflects the importance of the parameter $C$.

Equations (\ref {E:gp}) in principle cannot be used as they stand since they do not guarantee to give the best excitation frequencies and to satisfy the conservation laws. 
Indeed it is well known \cite {Burnet, boudj2011} that the inclusion of the anomalous average leads to a theory with a (unphysical) gap in the excitation spectrum. 
This can be seen to arise from the fact that the effective interaction between a pair of particles depends upon whether both come from the condensate or one is excited. 
The common way to circumvent this problem is to neglect $\tilde {m}$ in the above equations, which restores the symmetry and hence leads to a gapless theory. 
This is often reminiscent to the Popov approximation.
In addition, one finds that the anomalous average is divergent if one uses a contact interaction. To go beyond Popov one has to modify the atom-atom interaction.
This can be approximated, in the zero-momentum limit,  by a contact potential with position-dependent amplitude strength. Following \cite{Burnet, boudj 2013} we get:
\begin{align} \label{Ren}
&g_B |\Phi_B|^2\Phi_B+g_B\tilde{m}\Phi_B^{*}=g_B(1+\tilde {m}/\Phi_B ^2)|\Phi_B|^2\Phi_B 
\\=& U |\Phi_B|^2\Phi_B \nonumber.
\end{align} 
In a uniform gas, this definition of the effective potential $U(r)$ is related to the Beliaev-type second order coupling constant\cite {Bel, Griffin}. \\
It is useful to show that the new coupling constant (\ref {Ren}) is also equivalent to the many body $T$-matrix. 
As is known, this latter contains all the effects of the medium on the pair interactions in a gas, it is related to the low-momentum 
limit of the vacuum scattering amplitude $g_B$ via the following equation:
\begin{equation}  \label{Ren1}
T^{MB}=g_B \left\{1-T^{MB}\int \frac{d^3k}{(2\pi)^3}\left[\frac{2N_k+1}{2\epsilon_k}-\frac{1}{2E_k}\right]\right\},
\end{equation}
where $N_k=[\exp(\varepsilon_k/T)-1]^{-1}$ are occupation numbers for the excitations.  \\
The expression of $\tilde{m}$ which after the subtraction of the ultraviolet divergent part is given
\begin{equation}\label {Ren2}
\tilde{m}=- \Phi_B^2 T^{MB}\int \frac{d^3k} {(2\pi)^3} \left[\frac{2N_k+1}{2\epsilon_k}-\frac{1}{2E_k}\right].
\end{equation}
Comparison with (\ref{Ren1}), immediately yields the effective potential (\ref{Ren}).\\
Introducing now $U$ in (\ref {E:gp}), and using the fact that at very low temperature $2\tilde {n}+1\approx 2\tilde {m}$ \cite{boudj 2013}, one gets 
\begin{subequations}\label{E:td}
\begin{align}
&i\hbar \dot{\Phi}_B = \left [h_B^{sp}+g_B \left((\beta-2) |\Phi_B|^2+2n+\gamma|\Phi_I|^2\right) \right]\Phi_B, \label{E:td1} \\ 
&i\hbar \dot{\tilde{m}} = \left[ h_B^{sp}+g_B \left(2G \tilde {m}+2n+\gamma|\Phi_I|^2\right)\right]\tilde{m}, \label{E:td2}
\end{align}
\end{subequations}
where $\beta = U /g_B$, $G = \beta / 4(\beta-1 )$ and $\gamma=g_{IB}/g_B$.\\
Importantly, the set (\ref {E:td}) generates a self-consistent, gapless and non-divergent theory.\\
Let us now reveal the significance of parameter $\beta$. First of all, for $\beta = 1$, i.e., $ \tilde{m} = 0$, the new coupling constant $U$ reduces to $g_B$ 
and hence, Eq.(\ref {E:td1}) recovers the well known HFB-Popov equation. 
For $0<\beta<1$, $G$ is negative and hence, $\tilde{m} $ has a negative sign. 
For $\beta >1$, $G$ is positive, and thus, $\tilde{m} $ becomes a positive quantity.
For $\beta =2$, the gas becomes highly correlated and strongly interacting since $\tilde{m} =\Phi_B^2$.
Therefore, to guarantee the diluteness of the system, $\beta$ should vary as $\beta=1\pm\epsilon$ with $\epsilon$ being a small value.

To illustrate comparatively the behavior of the condensate, the anomalous density and the impurity density as a function of temperature, we consider that the mixture
is confined in an isotropic harmonic trap. We then solve numerically Eqs. (\ref {E:td})  and (\ref {E:gp2}) using the finite differences method in 3d model.\\  
In the numerical investigation,  we use $l_B=\sqrt{\hbar/ m_B\omega_{B}}$ and $\hbar \omega_{B}$ as the length (the ground state extent of a single BEC-boson particle) 
and the energy units, respectively, $\omega_{B} $ is the bosonic trapping frequency, and we end up with $\alpha=m_B/m_I$ being the ratio mass 
and $\Omega=\omega_{B}/\omega_{I}$, where $\omega_{I} $ is the impurity trapping frequency. 
The parameters are set to:  5\% of ${}^{85}$Rb impurity atom, $N$=$10^6$ of ${}^{23}$Na bosonic atoms, $a_B$=3.4nm, $a_{IB}=$ 16.7nm for repulsive interactions,  
$a_{IB}=-$16.7nm for attractive interactions and $\Omega=0.2$.

\begin{figure} 
\centerline{
\includegraphics[width=4.2cm,height=4cm]{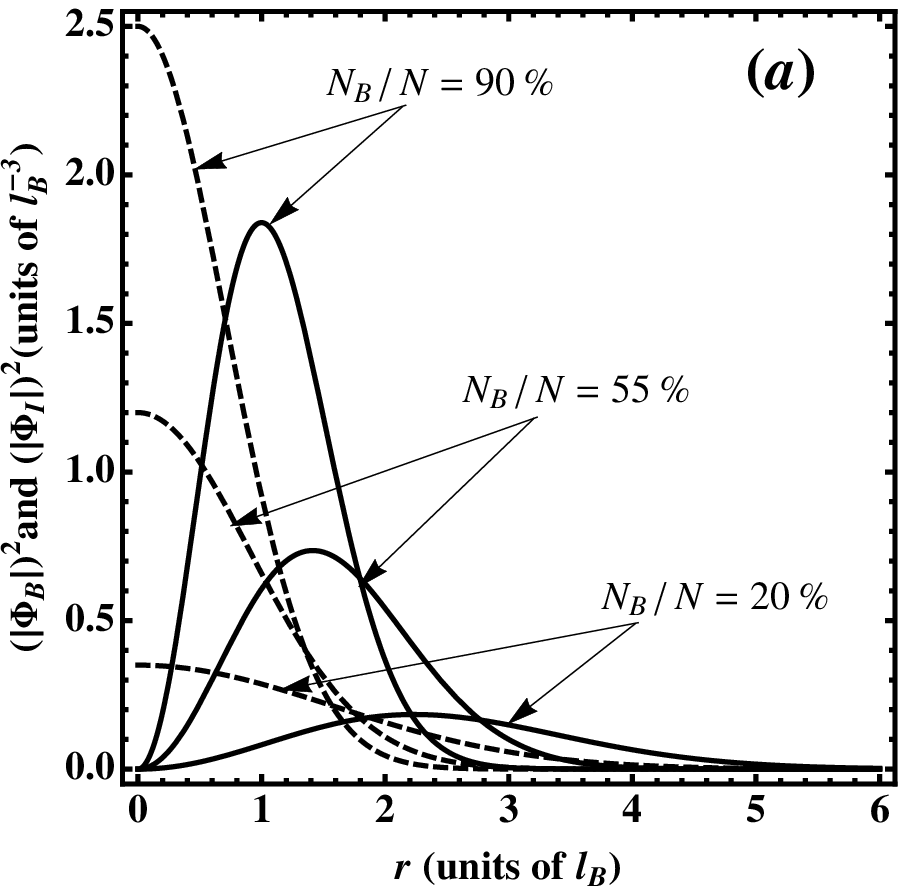}
\includegraphics[width=4cm,height=4cm]{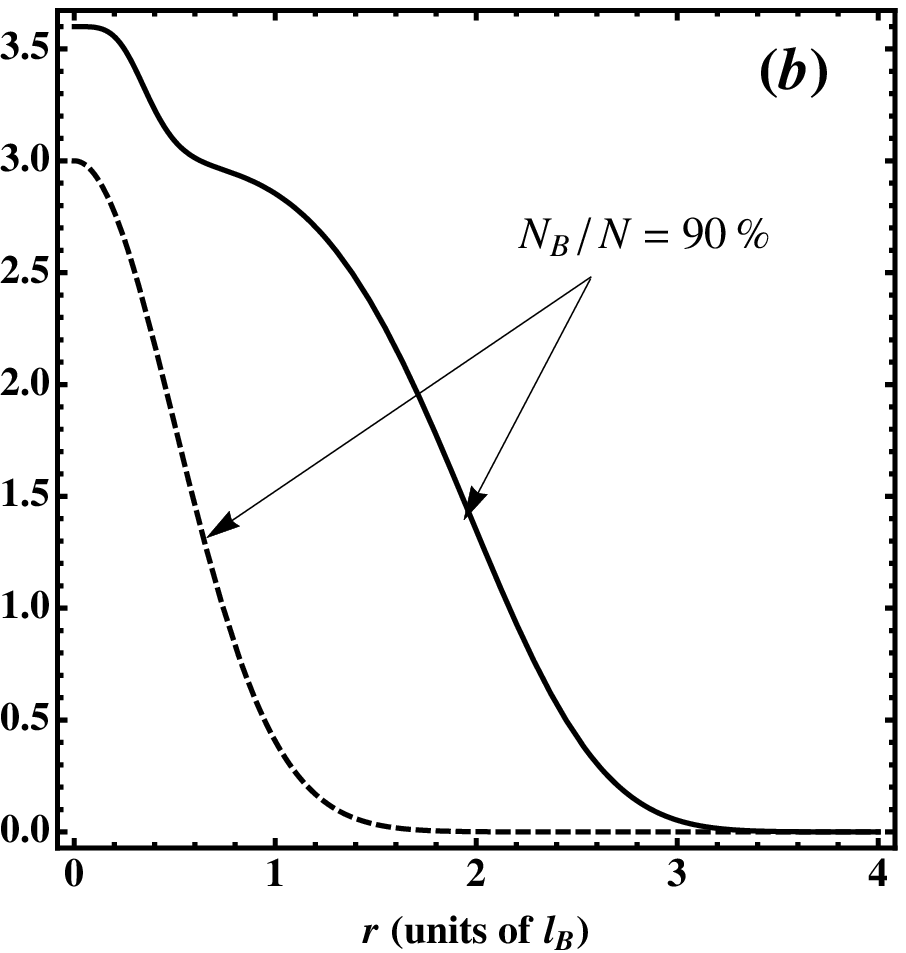}}
 \caption{Impurity (dashed lines) and condensed (solid lines) densities for various condensed fractions.
 (a) Repulsive interactions $a_{IB}$=16.7nm. (b) Attractive interactions $a_{IB}=-$16.7nm.}
\label{dens} 
\end{figure}

Figure.\ref {dens} (a)  depicts that for repulsive interactions, the condensate is distorted by the impurity and forms a dip near the center of the trap.
The impurity is focused inside the condensate forming a self-localized state in good agreement with existing theoretical results.
One can see also from the same figure that the density of the condensate is decreasing with temperature and the impurity becomes less localize (ignorant of its environment). 
The BEC exhibits a hump (peak) if the impurity and boson-atom mutually attract as is shown quantitatively in Fig.\ref {dens} (b). 
Indeed, the interaction potential $a_{IB}$ can be adjusted by means of the Feshbach resonance \cite {Piet}.\\
For $\beta = 1$, i.e. in the HFB-Popov approximation ($ \tilde{m} = 0$), the condensate and the impurity preserve their shape shown in Fig.\ref {dens} (a).
It has been pointed out that the inclusion of the anomalous density may shift the collective excitations of a trapped pure condensate rather than the shape of densities\cite{Burnet}.

For weak interactinons or small impurity concentrations, there is copious evidence that the impurity delocalizes and the BEC remains constant i.e. does not deform. 
This may lead to a coexistence of both quantities in the center of the trap.
For $a _{IB}=0$, the impurity and the condensate become completely decoupled.

\begin{figure} 
\centerline{
\includegraphics[scale=0.8]{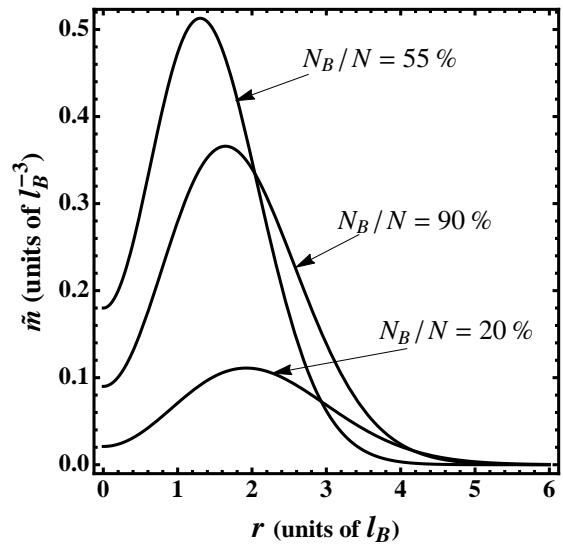}}
 \caption{Anomalous density for various condensed fractions and for $a_{IB}$=16.7nm.}
\label{dens1} 
\end{figure}

Figure.\ref {dens1}  shows that the anomalous density grows at low temperature until
it reaches its maximum value at intermediate temperatures and starts to disappear near the transition ($N_B/N\approx 55\%$) irrespective to the presence of the impurity or not.
The anomalous density is also distorted and formed a dip in an analogous manner with the condensate in the repulsive case. 

Let us now turn to discuss the relationship between our TDHFB equations (\ref{E:td}) and the HF-BdG equations.
Upon linearizing Eq.(\ref{E:td1}) around a static solution within the random phase approximation (RPA),  $\Phi_B=\Phi_{B0}+[u_k ({\bf r}) e^{-i\varepsilon_k t /\hbar}+v_k ({\bf r}) e^{i\varepsilon_k t/\hbar} ] e^{-i\mu_B t /\hbar}$. We then obtain
\begin{equation} \label{BdG}
\begin{pmatrix} 
{\cal L} & -{\cal M} \\
-{\cal M} & {\cal L}
\end{pmatrix}\begin{pmatrix} 
u_k({\bf r}) \\ v_k({\bf r})
\end{pmatrix}=\varepsilon_k \begin{pmatrix} 
u_k({\bf r}) \\ v_k({\bf r})
\end{pmatrix},
\end{equation} 
where ${\cal L}=(-\hbar^2/2m_B )\Delta + V_B+g_B \left(2 (\beta-2) |\Phi_B|^2+2n+\gamma|\Phi_I|^2\right)-\mu_B$, with 
$\mu_B$ being the chemical potential of the condensate component, it can be calculated from both Eq.(\ref{E:td1}) and the hydrodynamic approach (see section.\ref{Hydr}). 
 ${\cal M}=g_B (\beta-2)|\Phi_B|^2$ and $ u_k ({\bf r}), v_k ({\bf r})$ are the quasi-particle amplitudes.\\
The set (\ref{BdG}) constitutes the HF-BdG equations of BEC-impurity mixture at finite temperature.
Interestingly we remark that these equations are just the RPA of our Eq. (\ref{E:td1}). 

In the case of an impurity immersed in a homogeneous BEC, the excitation spectrum can be determined via the diagonalization of the whole RPA matrix of the set (\ref{E:gp}).
This yields an extended spectrum due to the condensate fluctuation corrections.
The obtained spectrum not only exhibits numerous quasiparticle properties of the attractive, repulsive and molecular branches as in the case of zero temperature \cite {Piet},
but permits us to examine effects of the condensate fluctuation on these branches. 


\section{Hydrodynamic approach and conservation laws} \label{Hydr}
Solving the TDHFB equations(\ref {E:td}) is not always the most adequate way to study the properties of BECs.
When it comes to the low-lying excitations, for example, it is often useful to switch to an equivalent treatment which is provided by the hydrodynamic equations.
These equations can be derived  by factorizing the condensate wave function and the anomalous density of the set (\ref {E:td})
according to the Madelung transformation \cite{boudj 2013}:

\begin{equation}  \label{eq9}
\begin{array}{rl}
&  \Phi_B({\bf r},t)=\sqrt{n_B({\bf r},t)} \exp (-i S({\bf r},t))
, \\ 

&\tilde{m}({\bf r},t)=\tilde{m}({\bf r},t) \exp (-i \theta({\bf r},t)),
\end{array}
\end{equation}
where $S$ and $\theta$ are phases of the order parameter and the
anomalous density, respectively. They are real quantities, related to the superfluid and thermal velocities, respectively, by
$v_B=\hbar/m_B {\bf \nabla} S$ and $v_{\tilde{m}}=\hbar/m_B {\bf \nabla} \theta$. By substituting expressions
(\ref {eq9}) in Eqs. (\ref {E:td})  and separating real and imaginary parts, one gets the following set of hydrodynamic equations:

\begin{equation}  \label{hydo1}
\frac{\partial n_B}{\partial t} +{\bf \nabla}.(n_B v_B)=0,
\end{equation}
\begin{equation}  \label{hydo11}
\frac{\partial |\tilde{m}|^2} {\partial t} +{\bf \nabla}.(|\tilde{m}|^2 v_{\tilde{m}})=0.
\end{equation}
Equations (\ref {hydo1}) and (\ref {hydo11}) are nothing else than equations of continuity
expressing the conservation of mass, and Euler-like equations read
\begin {widetext} 
\begin{equation}  \label{hydo2}
m_B\frac{\partial v_B}{\partial t} =-{\bf \nabla} \left [-\frac{\hbar^2}{2m_B} \frac{\Delta \sqrt{n_B}} {\sqrt{n_B}} +\frac{1}{2} m_B v_B^2 
+ V_B +g_B \left((\beta-2) n_B+2n+\gamma n_I \right) \right],
\end{equation}

\begin{equation}  \label{hydo22}
m_B\frac{\partial v_{\tilde{m}}}{\partial t} =-{\bf \nabla} \left [-\frac{\hbar^2}{2m_B} \frac{\Delta \tilde {m}} {\tilde {m}} +\frac{1}{2} m_B v_{\tilde{m}}^2
+ V_B+g_B \left(2G \tilde {m}+2n+\gamma n_I\right) \right],
\end{equation}
\end {widetext} 
where $\Delta \sqrt{n_B}/ \sqrt{n_B}$ and $\Delta\tilde {m}/\tilde {m}$ are, respectively, quantum and anomalous pressures, and
$n_I=|\Phi_I|^2$ is the density of the impurity.\\
In a nonstationary situation, we consider small oscillations (low density) for the condensed and anomalous densities around their static solutions in the form
\begin{equation} \label{hydo}
n_{B}=n_{B0}+\delta n_{B}, \qquad \tilde{m}=\tilde{m}_0+\delta\tilde{m}, 
\end{equation}
where  $\delta n_{B}/n_{B0}\ll1$ and $\delta\tilde{m}/\tilde{m}_0\ll1$. \\
Shifting the phases by $-\mu_B t/\hbar$ and $-\mu_{\tilde{m}} t/\hbar$, we then linearize Eqs.(\ref {hydo2}) and (\ref {hydo22}) with respect to $\delta n_{B}$, $\delta\tilde{m}$, ${\bf \nabla} S$ and ${\bf \nabla} \theta$, around the stationary solution.  The zero-order terms give
two expressions for the chemical potential:

\begin{equation}  \label{chimB}
\mu_B=-\frac{\hbar^2}{2m_B} \frac{\Delta \sqrt{n_B}} {\sqrt{n_B}}+ V_B +g_B \left((\beta-2) n_B+2n+\gamma n_I\right),
\end{equation}
and 
\begin{equation}  \label{chimm}
\mu_{\tilde{m}}=-\frac{\hbar^2}{2m_B} \frac{\Delta \tilde {m}} {\tilde {m}}+ V_B+g_B \left(2G\tilde {m}+2n+\gamma n_I\right),
\end{equation}
where $\mu_{\tilde{m}}$ is the chemical potential associated with the anomalous density. 
Strictly speaking  $\mu_{\tilde{m}}$ is also associated with the thermal cloud density since $\tilde{n}$ and $\tilde{m}$ are related to each other
by Eq. (\ref{eq6}).
Clearly $\mu_B\neq \mu_{\tilde{m}}$ at all ranges of temperature except near the transition where $n_B=\tilde{m}=0$ and $\tilde{n}=n$. 
Additionally, in the grand canonical ensemble the Hamiltonian may be written as $\hat K=\hat H-\mu \hat N$. If in the experiment only the total number
of particles $N = N_B +\tilde{N}$ or the total density $n$ can be fixed, then the total chemical potential of the system can be given as
\begin{equation}  \label{chimN}
\mu=\frac{N_B}{N} \mu_B+\frac{\tilde {N}}{N} \mu_{\tilde{m}},
\end{equation}
where $N_B/N$ and $\tilde {N}/N$ are, respectively, the condensed and the thermal fractions. It should be noted that this equation
arises naturally from our formalism without any subsidiary assumptions. Moreover, Eq. (\ref {chimN}) very nicely guarantees the
conservation of the total number of particles and precisely coincides with the theory of Ref \cite {Yuk} for pure BEC system.

We now focus on the important case of a spherically symmetric system in $d$-dimensions with the Bose gas trapped in the harmonic potential.
In the TF limit, it is reasonable to assume that the kinetic terms associated with both condensate and anomalous pressures can be neglected.
Therefore, Eqs (\ref{chimB}) and (\ref{chimm}) take the algebric form
\begin{equation}  \label{eq10}
n_B=\frac{1}{(\beta-2)} \left[\frac{\mu_B- m_B \omega_B^2 r^2 / 2}{g_B}-2n-\gamma n_I\right],
\end{equation}
\begin{equation}  \label{eq11}
\tilde {m}=\frac{2(\beta-1)}{\beta} \left[\frac{\mu_{\tilde {m}}-m_B \omega_B^2 r^2 / 2}{g_B}-2n-\gamma n_I\right].
\end{equation}
The density profiles (\ref {eq10}) and (\ref {eq11}) have the form of an inverted parabola. It is remarkable that for $\beta=1$, the anomalous density vanishes.\\
Then, using normalization conditions  $\int d^d r n({\bf r})=N$, $\int d^d r n_B ({\bf r})=N_B$, $\int d^d r n_I ({\bf r})=N_I$ and $\int d^d r\tilde {m}({\bf r})=N_{\tilde {m}}$,
Eqs. (\ref {eq10}) and (\ref {eq11}) provide useful formulas for the radius of the condensate and the anomalous density (i.e. the thermal cloud), respectively, as
\begin{equation}  \label{radB}
R_B=R_{TF}^{(0)}\left[(\beta-2)\frac{N_B}{N}+2+\gamma\frac{N_I}{N}\right]^{1/(d+2)},
\end{equation}
and 
\begin{equation}  \label{radM}
R_{\tilde{m}}=R_{TF}^{(0)}\left[\frac{\beta}{2(\beta-1)}\frac{N_{\tilde{m}}}{N}+2+\gamma\frac{N_I}{N}\right]^{1/(d+2)},
\end{equation}
where 
\begin{equation}  \label{radTF}
R_{TF}^{(0)}=\left[\frac{(d+2) \Gamma(\frac{d}{2}+1)l_B^2 N g_B}{\hbar \omega_B \pi^{d/2}}\right]^{1/(d+2)},
\end{equation}
is the $d$-dimensional TF radius at zero temperature and $\Gamma (x)$ is the Gamma function. 

At the classical turning point, one has $\mu=V_B (R)$. Then, using results of Eqs.(\ref {radB})  and (\ref {radM}), we obtain the following expressions 
for the chemical potential of the condensate and the anomalous density
\begin{equation}  \label{chimBB}
\mu_B=\mu_{TF}^{(0)}\left[(\beta-2)\frac{N_B}{N}+2+\gamma\frac{N_I}{N}\right]^{2/(d+2)},
\end{equation}
and 
\begin{equation}  \label{chimII}
\mu_{\tilde{m}}=\mu_{TF}^{(0)}\left[\frac{\beta}{2(\beta-1)}\frac{N_{\tilde{m}}}{N}+2+\gamma\frac{N_I}{N}\right]^{2/(d+2)},
\end{equation}
where 
\begin{equation}  \label{chimTF}
\mu_{TF}^{(0)}=\left[\frac{(d+2) \Gamma(\frac{d}{2}+1) }{2} Ng_B\left(\frac{\hbar \omega_B }{2\pi l_B^2}\right)^{d/2}\right]^{2/(d+2)},
\end{equation}
is the $d$-dimensional TF chemical potential at zero temperature. \\
Terms between brackets in the right hand side of expressions (\ref{radB}), (\ref{radM}), (\ref{chimBB}) and (\ref{chimII}) constitute the temperature and the impurity corrections to the radius 
and the chemical potential of both condensate and anomalous components. 

\begin{figure} 
\centerline{
\includegraphics[scale=0.8]{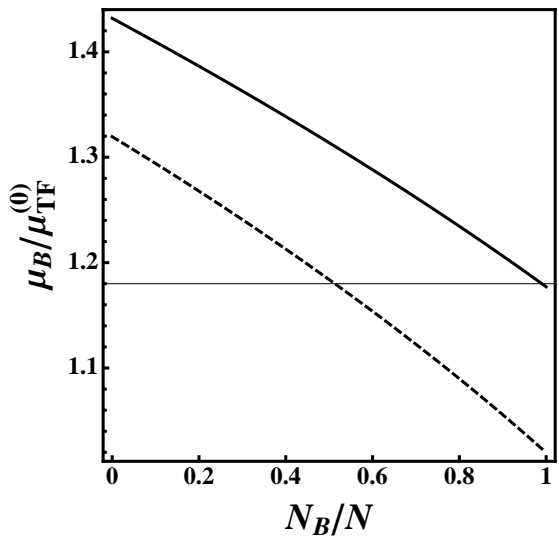}}
 \caption{Chemical potential of the condensate as a function of the condensed fracion with the same parameters as in Fig.\ref {dens} (a) for $d=3$.
Solid lines: in the presence of the impurity. Dashed lines : without impurity.}
\label{chim} 
\end{figure}

In 3d case, the behavior of the chemical potential of the condensate $\mu_B$ as a function of condensed fracion is
displayed in Fig.\ref{chim}. For repulsive interactions, $\mu_B$ increases when the condensed fraction decreases.
Moreover, a qualitative difference can be observed between $\mu_B$ with impurity and $\mu_B$ without impurity.
Fig.\ref{chim} clearly shows that the presence of the impurity in the system rises the chemical potential.

\begin{figure} 
\centerline{
\includegraphics[scale=0.8]{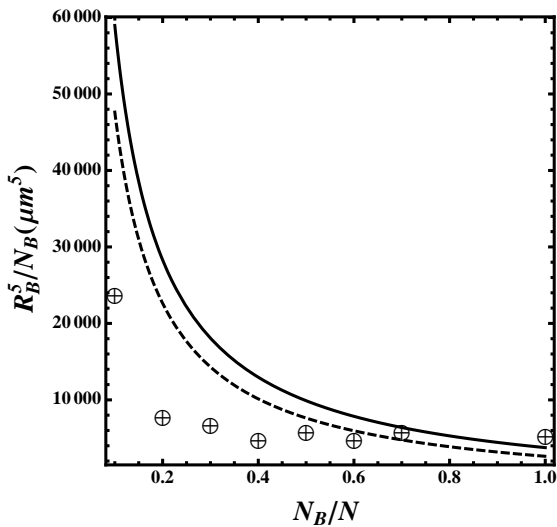}}
 \caption{The ratio $R_B^5/N_c$ as a function of the condensed fraction with the same parameters as in Fig.\ref {chim}.
Circles show experimental results of \cite {cara}. Solid lines : in the presence of the impurity. Dashed lines: without impurity.}
\label{rad} 
\end{figure}

Figure. \ref {rad} shows that in 3d case, $R_B^5/N_B$ increases when $N_B/N$ decreases and gives
reasonable agreement with experimental results of \cite {cara} for small values of $\beta$.
One can observe also that the impurity fraction tends to dilate the radius of the condensate.

Another important feature of the above generalized hydrodynamic equations 
is that they lead us to study in a straightforward manner the breathing modes of a BEC at nonzero temperatures in the presence of impurities. 
Inserting Eqs. (\ref{chimB}), (\ref{chimm}) into  (\ref{hydo2}) and (\ref{hydo22}) and taking the time derivative of the resulting equations, we find
\begin{equation}  \label{breth1}
 m_B\frac{\partial^2 \delta {n_{B0}}} {\partial t^2} ={\bf \nabla}. (n_{B0} {\bf \nabla} \mu_B),
\end{equation}
and
\begin{equation}  \label{breth2}
 m_B\frac{\partial^2 \delta \tilde{m}_0} {\partial t^2} ={\bf \nabla}. (\tilde{m}_0 {\bf \nabla} \mu_{\tilde{m}}).
\end{equation}
Equations (\ref {breth1}) and (\ref {breth2}) describe the collective modes of the condensate
and the anomalous density for a Bose gas in an arbitrary potential and in the presence of an impurity. 
The validity of these equations is based on the assumption that the spatial variation of densities ($n_{B0}$ and $\tilde{m}_0$) are smooth not only in the ground state, but also during the oscillation.
In a uniform gas, this is equivalent to requiring that the collective frequencies be much smaller than the chemical potential.
For $\gamma=0$ and at very low temperature where $n_B/\tilde{m}\ll 0$, Eq.(\ref {breth1})  well recovers the famous Stringari's equation
$m \partial^2 \delta {n_{0}}/ \partial t^2 ={\bf \nabla}. (n_{0}{\bf \nabla} \mu)$ \cite{String} which describes the collective modes of a pure condensate
at zero temperature.

\section{Conclusion and outlook} \label{concl}
In this paper we have derived  a set of coupled equations for BEC-impurity mixture using the BV time-dependent variational principle. 
These equations govern in a self-consistent way the dynamics of the condensate, the thermal cloud, 
the anomalous average and the impurity at finite temperature. We have shown that the TDHFB equations satisfy all the conservation laws and provide a gapless excitations spectrum.
The numerical simulations of the TDHFB equations in an isotropic harmonic trap showed that the condensate and the anomalous density 
are deformed by the impurity for repulsive interactions. While in the attractive case, the condensate forms a hump  near the impurity. 
In addition, the impurity becomes ignorant of its host medium at higher temperatures.

Moreover, we have derived quite useful analytic expressions for the chemical potential and for the radius of both condensate 
and of anomalous components in the presence of the impurity using the TF approximation in $d$-dimensional model.  
The obtained expressions appear as natural extensions of those existing in the literature since they gather both temperature and impurity corrections. 
We have found in this sense that the impurity may strongly enhance the chemical potential and the radius of the mixture. 
Importantly, these expressions can be used to investigate the expansion, the breathing modes and the transport properties of BEC-impurity mixtures at nonzero temperatures.


Finally, it would be interesting to discuss the importance of the three-body forces in the impurity-host-host mixtures. These effects are indeed ubiquitous and arise naturally in effective field theories when one integrates out some of high-energy degrees of freedom in the system \cite {Ham}.
In the spirit of the BV variational principle, three-body effects can be taken into account by using a post-Gaussian ansatz \cite{Flu} i.e. adding the three-boson operator 
$W_{i,j,k} \alpha_i \alpha_j \alpha_k/3$ to the expressions (\ref {eq1}) and (\ref{eq2}), where the matrix $W_{i,j,k}$ is symmetrical with respect to its three indices.
This leads to extend the TDHFB equations (Eq.(\ref{E:gp})) and provide a self-consistent equation of motion for the triplet correlation function.
However, the Efimov effect \cite {Efi} happens in such a system if the impurity-host interaction is resonant even if the host-host interaction is not. 
The effect is due to an effective $1/R^2$ ($R$ is the hyper radius) attraction in this three-body system and it gets enhanced when the impurity is light. 
In this situation, one can expect that there is no suppression of the three-body relaxation  \cite {Petrov} and thus, the polaron cannot survive for a long time in the degenerate regime.
 
\section{Acknowledgements}
We are grateful to Marcel V\'en\'eroni for careful reading of the paper and for helpful discussions. We also
acknowledge stimulating discussions with D. S. Petrov.
We would like to thank the LPTMS, Paris-sud for a visit, during which part of this work has been done.

\end{document}